\documentstyle[12pt]{article}
\begin{document}
\title{The Two-fluid Description of a Mesoscopic Cylinder}

\author{M. Stebelski, M. Lisowski and E. Zipper\\
        Institute of Physics, University of Silesia\\
        ul. Uniwersytecka 4, 40 - 007 Katowice, Poland}

\baselineskip=15pt

\maketitle

\begin{abstract}

 \vspace{0.5cm}

\end{abstract}

\newpage
\section{Introduction}

   The transport properties of mesoscopic metallic or semiconducting samples at
low temperature have been shown to exhibit features characteristic of the
quantum coherence of the electronic wave function along the whole sample, see
e.g. \cite{Washburn}. It is thus interesting to compare the characteristics of 
mesoscopic systems with those for other coherent systems.

   The aim of the present work is to show that if we reduce the dimensions of
a cylinder made of a normal metal or a semiconductor to mesoscopic dimensions
and if we assume that electrons interact via the magnetostatic interaction, a
system exhibits coherent properties absent in macroscopic samples.

   Coherent systems like superfluids or superconductors can be in general
described by a two-fluid model namely the total density $n$ is equal to the sum
of the densities of the coherent $n_{c}$ and normal $n_{n}$ components

    \begin{equation}
    n=n_{c}+n_{n}.
    \end{equation}
For macroscopic, nonsuperconducting systems (here after refered as "normal") 
$n_{c}=0$.    

   In this paper we discuss the electromagnetic response and its static limit 
of a mesoscopic cylinder of very small thickness $d$ made of a normal metal or
semiconductor. The circumference and height of the cylinder are denoted by $L_
{x}, L_{y}$ respectively $(d \ll L_{x}, L_{y})$.

   We argue that such mesoscopic cylinder can be in general described by the 
two-fluid model. The coherent state encountered here, although in principle
different from the superconducting state, exhibits the following properties 
which bear resemblance to superconductors:

\begin{itemize}

\item free acceleration of part of electrons

\item strong diamagnetic susceptibility at low fields due to the reduction of
paramagnetic susceptibility

\item dynamic energy gap coming from magnetostatic coupling of electrons

\item persistent currents in the presence of the static flux

\item flux trapping

\end{itemize}

   The relation between the induced current density $J(\underline{q}, w)$
and the electric field $E(\underline{q}, w)$ is generally written as

    \begin{equation}
    \label{pierwsze}
    J(\underline{q}, w) = \frac{iK(\underline{q},w)}{
    w} E(\underline{q},w).
    \end{equation}
It is the form of the kernel $K(\underline{q}, w)$ that determines the 
properties of the system both static as well as for finite frequencies. The
following limits imply

    \begin{eqnarray}
    \label{granica1}
    \lim_{w \rightarrow 0} \lim_{\underline{q} \rightarrow 0} K(
    \underline{q}, w)  \equiv - \chi \neq 0 & \mbox{- infinite 
    conductivity}, \\
    \label{granica2}
    \lim_{\underline{q} \rightarrow 0} \lim_{w \rightarrow 0} K(
    \underline{q}, w)  \neq 0 & \mbox{- Meissner effect .} 
    \end{eqnarray}
For normal electrons in macroscopic metallic or semiconducting samples both 
limits are zero.

   In this paper we calculate the conductivity $\sigma(w)$ and we study the 
first limit only as we are dealing with a narrow cylinder and we neglect the 
$q$ dependence of the vector potential $\bf{A}$. The results for a thick 
cylinder will be presented in a forthcoming paper.

   We show that the relation (\ref{granica1}) is fulfilled in the 
nonsuperconducting mesoscopic cylinder and its magnitude depends on the shape 
of the Fermi Surface (FS).

   The reaction of a one-dimensional mesoscopic ring to a static and time 
varying magnetic field and coupled to a thermal bath has been investigated by 
Trivedi et. al. \cite{Trivedi}. They discussed using the Kubo linear 
response theory the properties of the dynamic and static response function. 
Such a one channel system is a simplification - experimental samples have many 
transverse channels and the question arises how do the system characteristics 
depend on the transverse dimensions.
 
    We extend the calculations of the conductivity of a one-dimensional 
disordered ring \cite{Trivedi} driven by the electromagnetic field to the 
system of cylinder geometry and check whether coherence will still be 
maintained in a system with many transverse channels. 

   We also discuss the orbital magnetic susceptibility and show that it 
exhibits anomalous diamagnetism related to the presence of a dynamic gap.

   We further investigate persistent currents which arise as a result of the 
flux sensitivity of energy levels. We discuss coherence of currents from 
different channels and the magnitude of persistent currents for different 
shapes of the FS. 

   In one of our recent papers \cite{Stebelski} we discussed persistent, 
paramagnetic currents in the free electron model with different shapes of the 
FS. The FS was assumed to be a function of certain parameters which described a
change of its shape from circular to nearly quadratic. In this paper we use the
tight-binding model where different shapes of the FS follow in a natural way 
from the crystal structure and band filling \cite{Anselm}. We discuss here, 
contrary to \cite{Stebelski}, diamagnetic persistent currents.

   Finally we address the issue of whether a system of interacting electrons 
confined to a mesoscopic cylinder can sustain a spontaneous persistent 
currents in the absence of an externally applied flux. These currents 
correspond to flux trapping in the cylinder.

\section{Coherent properties of a mesoscopic cylinder}
   
   Let us consider a system of spinless electrons constrained to move on a 
hollow cylinder with $M$ channels in the height $L_{y}$ and $N$ sites with 
lattice spacing $a$ in the circumference $L_{x}$ threated by the magnetic flux
$\phi$. We assume that electrons interact via the long range magnetostatic
(curent-curent) interaction, the interaction is taken here in the mean field
approximation (MFA). The tight binding Hamiltonian is of the form   

    \begin{eqnarray}
    \label{hamiltonian}
    {\cal H} = \sum_{n m} [(2t + V_{n m})c_{n m}^{+}c_{n m} - t e^{i\Theta_{n 
    n+1}} c_{n+1 m}^{+}c_{n m} + \nonumber\\
    - te^{-i\Theta_{n n+1}}c_{n m}^{+}c_{n+1 m} - tc_{n m+1}^{+}c_{n m} - tc_
    {n m}^{+}c_{n m+1}],
    \end{eqnarray}
where $t=\frac{\hbar^{2}}{2m_{e}a^{2}}$, $\Theta$ comes from the magnetic flux 
$\phi$ 

    \begin{equation}
    \Theta_{n n+1}=\frac{eaA}{\hbar c}=\frac{ea\phi}{\hbar cL_{x}}=\frac{e\phi}
    {\hbar cN}, \; \; \phi=L_{x}A,
    \end{equation}
A is the vector potential in the $x$ direction. The flux $\phi$ is composed of 
two parts 

    \begin{equation}
    \label{strumien}
    \phi = \phi_{e} + \phi_{I},\;\;\phi_{I}={\cal L} I(\phi),
    \end{equation}
i.e. each electron moves in the external magnetic flux $\phi_{e}$ and in the 
flux coming from the current $I$ in the cylinder, ${\cal L}$ is the 
selfinductance coefficient. The second term in equation (\ref{strumien}) comes 
from the magnetostatic coupling.

   For a clean sample ($V_{n m}=0$) we can diagonalize the Hamiltonian directly
${\cal H} \equiv {\cal H}^{xy} = {\cal H}_{x}
+{\cal H}_{y}$

    \begin{equation}
    {\cal H}^{x y} = \sum_{k_{x}} \sum_{k_{y}} 2t\left[1-\cos\left( k_{x}a - 
    \frac{e\phi}{\hbar cN}\right) - \cos(k_{y}a)\right]n_{k_{x}k_{y}},
    \end{equation}
    \begin{equation}
    k_{x}=k_{x}(s,\phi)=\frac{2\pi}{Na}\left(s+\frac{\phi}{\phi_{0}}\right), \;
    \; s = 0, \pm 1,\ldots,
    \end{equation}
    \begin{equation}
    k_{y}=k_{y}(r)=\frac{r\pi}{(M+1)a}, \; \;r = 1, \ldots M,
    \end{equation}
    \begin{equation}
    \psi_{sr}(x,y) = \sqrt{\frac{2}{N(M+1)}}e^{ik_{x}x} \sin(k_{y}y),
    \end{equation}
    \begin{equation}
    {\cal H}^{x y} \mid sr \rangle = {\cal E}_{sr} \mid sr \rangle,
    \end{equation}
    \begin{equation}
    \label{energia}
    {\cal E}_{sr}= 2t\left[1- \cos\left(\frac{2\pi}{N}\left(s-\frac{\phi}
    {\phi_{0}}\right) \right) - \cos \left(\frac{r \pi}{M+1}\right)\right] 
    \equiv {\cal E}_{s}+{\cal E}_{r}.
    \end{equation}
The eigenstates of equation (\ref{hamiltonian}) we denote by $\mid \alpha 
\rangle$, these are the eigenstates of the $\cal H$ with impurities which can 
be diagonalized by computer calculations. Here we can treat the effects of
disorder in rough analogy to the effects of temperature assuming that the 
elastic scattering causes transitions between the states of the perfect system 
\cite{Riedel}. We can expand the eigenstates of disordered cylinder $\mid 
\alpha \rangle$ in terms of those of the perfect cylinder $\mid sr \rangle$,

    \begin{equation}
    \mid \alpha \rangle = \sum_{sr} a_{sr}^{\alpha} \mid sr \rangle, \; \; a_
    {sr}^{\alpha} = \langle sr \mid \alpha \rangle,
    \end{equation}
    \begin{equation}
    {\cal H} \mid \alpha \rangle = {\cal E}_{\alpha} \mid \alpha \rangle, \; \;
    \rho_{0} = \frac{1}{\exp [\beta ({\cal H} - \mu)]+1},\;\; \rho_{0}\mid
    \alpha\rangle = f_{\alpha} \mid \alpha \rangle,
    \end{equation}
where $f_{\alpha} = \frac{1}{\exp[{\beta({\cal E}_{\alpha}- \mu)]}+1}$ is the 
FD distribution function, $\mu$ is the chemical potential. 

   In this paper we consider the case where the number of electrons $N_t$ is 
kept fixed and the chemical potential depends on flux; $\mu \equiv \mu 
(N_t, \phi)$ can be calculated from the equation $\sum_{\alpha}f_{\alpha}= 
N_t$.

   To introduce relaxation we assume a small coupling exists between the 
cylinder and the thermal bath, with the relaxation rate denoted by $\gamma$.
Then following \cite{Trivedi}, by use of the kinetic equation for the density 
matrix $\hat{\rho}(t)$ we calculate the response of the system in the 
relaxation time approximation. For details of the formalism see \cite{Trivedi}.

   Now we apply a small time dependent flux $\delta \phi = L_{x} \delta A$,
where $\delta A$ has only the tangential component. Denoting by $\delta {\cal 
H}$ the corresponding change in the Hamiltonian we can write

    \begin{equation}
    \delta {\cal H}\equiv-\frac{\delta \phi}{c} \hat{I_{i}}= -\frac{\delta 
    \phi}{c} \hat{I_{p}}-\frac{1}{2}\left(\frac{\delta \phi}{c}\right)^{2}\hat
    {D},
    \end{equation}
where $\hat{I_{i}}=\hat{I_{p}}+\hat{I_{d}}$, $\hat{I_{i}}$ is the induced 
current operator.

    \begin{equation}
    \hat{I_{p}} = i\frac{e \hbar}{2m_{e}aL_{x}} \sum_{nm} (e^{i \Theta_{n n+1}
    }c_{n+1 m}^{+}c_{nm} + h.c.).
    \end{equation}
$\hat{I_{p}}$ is the operator of paramagnetic current, running along the 
cylinder circumference, $\hat{I_{d}}$ is the diamagnetic current operator 

    \begin{equation}
    \hat{I_{d}}=-\frac{\delta \phi}{cL_{x}^{2}} \hat{D},\;\; \hat{D}=\frac{e^
    {2}}{2m_{e}} \sum_{nm} (e^{i \Theta_{n n+1}}c_{n+1 m}^{+}c_{nm} + h.c.).
    \end{equation}
Calculating by the second order perturbation theory the changes in the energy 
caused by $\delta {\cal H}$ and comparing them with the Taylor expansion of
the energy we obtain important relations

    \begin{equation}
    \label{pa}
    \langle \alpha \mid \hat{I_{p}} \mid \alpha \rangle = - c \frac{\partial 
    {\cal E}_{\alpha}}{\partial \phi},
    \end{equation}
    \begin{equation}
    \label{dia}
    \frac{1}{L_{x}^{2}}\langle \alpha \mid \hat{D}\mid \alpha \rangle + 2 \sum_
    {\beta}\frac{\mid \langle \alpha \mid \hat{I_{p}} \mid \beta \rangle \mid^
    {2}}{{\cal E}_{\alpha} - {\cal E}_{\beta}} = c^{2} \frac{\partial^{2}{\cal 
    E}_{\alpha}}{\partial \phi^{2}}.
    \end{equation}
Calculating the current induced by the small time dependent flux $I_{i}=\mbox
{Tr} \hat{I_{i}}\hat{\rho}$ we can find the conductivity $\sigma(w)$

    \begin{equation}
    \label{gestosc}
    J(w)=\frac{L_{x}}{V}I_{i}(w)=\sigma(w)E(w),
    \end{equation}
where

    \begin{equation}
    E(w)=\frac{iw}{cL_{x}} \: \delta \phi.
    \end{equation}
Making use of (\ref{pa}) and (\ref{dia}) we get after some algebra

    \begin{equation}
    \label{sigma}
    \sigma(w) = \frac{L_{x}^{2}}{V}\left[-\frac{ic}{w} \frac{\partial I}
    {\partial \phi}- \frac{c^{2}}{\gamma-iw} \sum_{\alpha}\frac
    {\partial f_{\alpha}}{\partial \phi} \frac{\partial {\cal E}_{\alpha}}
    {\partial \phi} + i {\sum_{\alpha \beta}}^{'}\frac{f_{\alpha}-f_{\beta}}
    {{\cal E}_{\alpha}-{\cal E}_{\beta}} \frac{\mid \langle \alpha \mid \hat{I}
    _{p}\mid \beta \rangle \mid^{2}}{{\cal E}_{\alpha}-{\cal E}_{\beta}- w - i
    \gamma}\right],
    \end{equation}
where $I$ is the equilibrium persistent current

    \begin{equation}
    \label{prad1}
    I=\mbox{Tr}\rho_{0}\hat{I_{p}}=-c\sum_{\alpha} f_{\alpha}\frac{\partial
    {\cal E}_{\alpha}}{\partial \phi},
    \end{equation}
V denotes the volume of the sample. 

   Up to now the formal calculations went in an analogous way as in 
\cite{Trivedi}, the new physics appears as we perform the summation about the 
transverse channels. The first two terms in (\ref{sigma}) are present only in 
mesoscopic multiply connected structures where the energy levels are flux 
sensitive. The first term, being purely imaginary, determines coherent response
of the system. The second term involves the intralevel scattering and depends 
essentially on $\gamma$, the third term involves interlevel transitions and is 
connected with the elastic scattering. In the absence of the elastic scattering
$\hat{I_{p}}$ is diagonal in the unperturbated basis $\mid sr \rangle$ and the 
third term is zero. The ac conductivity in a multichannel metallic ring in the 
diffusive regime has been extensively discussed in \cite{Reulet}.

   In this paper we will focus our attention on the reactive response in the 
static limit. We want to establish the conditions under which the system 
exhibits strong coherent behaviour.

   Calculating the $w=0$ limit of equation (\ref{sigma}) we find

    \begin{equation}
    \label{Im}
    \lim_{w \rightarrow 0} w \: \mbox{Im} \, \sigma (w) = - \frac{c}{V}
    \frac{\partial I}{\partial \phi}.
    \end{equation}
Thus the conductivity exhibits an imaginary part which in the low frequency 
limit is proportional to the flux derivative of persistent current.

   It follows from the basic theory of conductivity \cite{Ashcroft} that 
coherent electrons i.e. those which run without scattering obey the following 
relation

    \begin{equation}
    \label{przed}
    \lim_{w \rightarrow 0} w \, \mbox{Im} \, \sigma (w) = \frac{n_{c}e^{2}}{m_
    {e}},
    \end{equation}
where $n_{c}$ is the density of coherent electrons. From equations 
(\ref{pierwsze}), (\ref{granica1}), (\ref{gestosc}) and (\ref{Im}) we find

    \begin{equation}
    \label{ost}
    \lim_{w \rightarrow 0} \lim_{\underline{q} \rightarrow 0} K(q,w) \equiv - 
    \chi = - \frac{c}{V} \frac{\partial I}{\partial \phi},
    \end{equation}
or from (\ref{przed})

    \begin{equation}
    \label{chi}
    \chi=-\frac{n_{c}e^{2}}{m_{e}}.
    \end{equation}
The finite limit of the kernel $K(\underline{q},w)$ corresponds to free 
acceleration of the part $(\equiv n_{c})$ of electrons - the feature 
characteristic of superconductors.

   Using the relation between $\sigma(w)$ and $\chi(w)$ \cite{Pines}

    \begin{equation}
    \label{11.1}
    \sigma(w)=-\frac {i}{w} \chi (w)
    \end{equation}
we can calculate the static orbital susceptibility

    \begin{equation}
    \label{11.2}
    \chi =\lim_{w \rightarrow 0}\chi (w)=\lim_{w \rightarrow 0}iw \sigma (w)=-
    \lim_{w \rightarrow 0}w \mbox{Im}\sigma (w )=-\frac{n_ce^2}{m_{e}},
    \end{equation}
where relation (\ref{przed}) has been used. 

   We would like to stress that the equilibrium susceptibility and the zero
frequency limit of a dynamic response coincide \cite{Jansen} as we are working
with the constant number of particles. From equations (\ref{dia}),
(\ref{prad1}) and (\ref{ost}) we obtain 

    \begin{equation}
    \label{11.3}
    \chi =\chi_{D}+\chi_{p},
    \end{equation}
where

    \begin{equation}
    \label{11.3.5}
    \chi_{D}=-\frac{1}{V}\sum_{\alpha}f_{\alpha} \langle \alpha\mid\hat{D}\mid
    \alpha \rangle =-\frac{e^2}{m_{e}}n ,
    \end{equation}
$\chi _D$ is the diamagnetic susceptibility, $n$ is the electron density

    \begin{equation}
    \label{11.4}
    \chi_{p}=-\frac{L_{x}^{2}}{V}\left[ 2\sum_{\alpha\beta}f_{\alpha}\frac{\mid
    \langle \alpha \mid \hat{I}_{p}\mid\beta \rangle \mid^{2}}{{\cal E}_{\alpha
    }-{\cal E}_{\beta}}+c^{2}\sum_{\alpha}\frac{\partial f_{\alpha}}{\partial 
    \phi}\frac{\partial {\cal E}_{\alpha}}{\partial \phi}\right] 
    \end{equation}
$\chi_{p}$ is the paramagnetic susceptibility.

   Equations (\ref{11.2}) - (\ref{11.3.5}) give the following relation for 
$\chi_{p}$

    \begin{equation}
    \chi_{p}=\frac{n_{n}e^{2}}{m_{e}},\;\; n_{n}=n-n_{c}.
    \end{equation}
We see that $n_{n}$ is the density of normal electrons which undergo different
kinds of scattering processes. Both terms in $\chi_{p}$ depend on the shape of
the FS.

   For a macroscopic "normal" sample the distance between energy levels is 
small and the occupation probability is then a smooth function of the energy.
If is then 
justified to replace the sum in equation (\ref{prad1})  by an integral what 
leads to vanishing currents and to zero limit in equation (\ref{ost}). Thus 
in macroscopic samples $\chi_p$ and $\chi_D$ nearly cancel each other and the
system displays only a small residual diamagnetism.

   However the mesoscopic cylinder with $L_{x} \sim 1 \mu m$ has large energy 
spacings in $x$ direction and the occupation probability near FS changes 
significantly over these spacings. The sum can not be then replaced by the 
integral and we get a finite current $I$ which is persistent at low T because 
of Quantum Size Energy Gap (QSEG) \cite{Cheung, Szopa}. The paramagnetic 
susceptibility is
drastically reduced due to the presence of the energy gaps and the system 
displays an anomalous diamagnetism. This problem will be further discussed in 
chapter 3. We see that in mesoscopic system the paramagnetic and diamagnetic 
susceptibilities fail to cancel ($n \neq n_{n}$) and we can describe the 
system by the two-fluid model.

   The density of coherent electrons $n_{c}$ is proportional to the flux 
derivative of persistent current.

   We shall discuss now the current $I$ in more detail. It's magnitude depends
on the strength of the phase correlations between currents of different 
channels. 

   A large phase correlation among channel currents means that the increase of 
the flux $\phi$ results in an almost simultaneous cross of the FS by the large 
number of channels. The most favorable situation takes place if the separation 
between the last occupied level and the FS from channel to channel is nearly 
the same. There exists then a perfect correlation among the channel currents 
because the $M$ levels cross the FS simultaneously while the flux is changed by
one fluxoid - we get then the largest amplitude of the total current.

   Let us assume that our mesoscopic cylinder contains a small number of
impurities. The average current, where the average is taken over impurity
configurations, can be calculated in the linear response approach. For the $M$
channel system one obtains \cite{Entin} 

    \begin{equation}
    \label{7.4}
    I(\phi )\approx \exp\left(-\frac{L_{x}}{2\lambda}\right)\sum_{r=1}^{M}\sum_
    {s=0,\pm 1}^{\pm \infty}f_{sr}I_{s},
    \end{equation}
where 

    \begin{equation}
    I_{s}=\frac{e\hbar}{m_{e}aL_{x}} \sin \frac{2\pi}{N}(s-\phi^{'})\;\;\;
    s=0,\pm 1,\ldots\;\; \phi ^{'}=\frac{\phi}{\phi_{0}},\;\; \phi_{0}=\frac
    {hc}{e}, 
    \end{equation}
$\lambda$ is the mean free path. For a sufficiently clean material $\lambda$ 
can be of the order of a few microns and the impurities do not decrease
the current significantly.

   The current $I(\phi)$ is periodic in $\phi^{'}$ with period 1 and can be 
expressed as a Fourier sum \cite{Cheung}

    \begin{equation}
    \label{prad}
    I(\phi) = \exp \left(-\frac{L_{x}}{2 \lambda}\right) \sum_{m=1}^{M}
    \sum_{l=1}^{\infty} \frac{4 T}{\pi T^{*}} \frac{2et}{N \hbar} \frac
    {\exp\left(-\frac{lT}{T^{*}}\right)}{1-\exp\left(-\frac{2lT}{T^{*}}\right)}
    \sin \left(\frac{2 \pi l \phi}{\phi_{0}} \right)F(m),
    \end{equation}
where  $\lambda$ is the mean free path, $T^{*}$ is characteristic temperature 
set by the level spacing at the Fermi surface for an electron moving in the $x$
direction,

    \begin{equation}
    \label{Fermi}
    F(m)=\sin (k_{F_{x}}(m)a) \cos(lNk_{F_{x}}(m)a),
    \end{equation}
$F(m)$ is the factor depending on a shape of the FS, $k_{F_{x}}$ is calculated 
from the equation for the FS. 

   In the tight-binding approximation the shape of the FS depends on the 
lattice type and on the filling factor \cite{Anselm}.

   For a 2D square lattice for the half-filled band the equation for the FS 
reads

    \begin{equation}
    \cos(ak_{F_{x}})+\cos(ak_{F_{y}})=0
    \end{equation}
and the equation for $F(m)$ takes the form

    \begin{equation}
    F(m)= \sin (\arccos(-\cos (ak_{F_{y}}(m)))) \cos(lN\arccos(-\cos(ak_{F_{y}}
    (m)))).
    \end{equation}
The FS is than quadratic with the diagonals along $k_{x}, k_{y}$ axes. 

   For the filling factor much less than a half we can expand the cosines in 
the dispersion relation (\ref{energia}) for small $k$-values and we obtain

    \begin{equation}
    \label{nowae}
    {\cal E}_{sr}= 2t\left[\frac{1}{2} \left(\frac{2\pi}{N}\left(s-\frac{\phi}
    {\phi_{0}}\right) \right)^{2}-\frac{1}{2}\left(\frac{r\pi}{M+1}\right)^{2}
    -1\right] \equiv {\cal E}_{s}+{\cal E}_{r}.
    \end{equation}
The FS is then circular and equation (\ref{Fermi}) reads

    \begin{equation}
    F(m)=\sin\left(2\sqrt{1-\left(\frac{ak_{F_{y}}(m)}{2}\right)^{2}}\right)  
    \cos\left(2lN\sqrt{1-\left(\frac{ak_{F_{y}}(m)}{2}\right)^{2}}\right).
    \end{equation}
We can also imagine a cylinder made of a set quasi 1D mesoscopic rings stacked
along $y$ axis. The FS will be then nearly flat perpendicular to $k_{x}$ 
direction and the formula for the current is then of the form

    \begin{eqnarray}
    I(\phi) = \exp \left(-\frac{L_{x}}{2 \lambda}\right) M \sum_{l=1}^
    {\infty} \frac{4 T}{\pi T^{*}} \frac{2et}{N\hbar} \frac{\exp\left(-
    \frac{lT}{T^{*}}\right)}{1-\exp\left(-\frac{2lT}{T^{*}}\right)}\sin \left(
    \frac{2 \pi l \phi}{\phi_{0}} \right) \times \nonumber \\
    \times \sin(k_{F_{x}}a)\cos(lNk_{F_{x}}a).
    \end{eqnarray}
   In order to avoid perfect nesting in all figures we present the results for 
slightly less than half filled bands instead of half filled ones. The Fermi 
Surfaces for such cases will have then rounded corners.

   In Fig.1 \marginpar{Insert Fig.1} we present persistent currents for four
different shapes of 2D FS (inserted figure). We see that the amplitude of the 
current increases with increasing departure of the FS from circular because the
phase correlation of the channel currents increases. 

   We have found that for a 2D squared FS, drawn by a solid line in the 
inserted figure, maximal interchannel phase correlations exists for $\frac
{L_{x}}{L_{y}}=2+\frac{n}{3}$, where $n$ is a positive integer. For other 
values of the ratio $\frac{L_{x}}{L_{y}}$ the current is much weaker.

   The FS of the shapes as in the inserted figure are frequently met in High 
$T_{c}$ Superconductors with 2D conduction \cite{Ruvalds,King}.

   For a 3D case with $P$ channels in the cylinder thickness the current 
(\ref{prad1}) takes a form

    \begin{equation}
    \label{prad3d}
    I(\phi) = \exp \left(-\frac{L_{x}}{2 \lambda}\right) \sum_{p=1}^{P}
    \sum_{m=1}^{M}\sum_{l=1}^{\infty} \frac{4 T}{\pi T^{*}} \frac{2et}{N \hbar}
    \frac{\exp\left(-\frac{lT}{T^{*}}\right)}{1-\exp\left(-\frac{2lT}
    {T^{*}}\right)}\sin \left(\frac{2 \pi l \phi}{\phi_{0}} \right)F(m,p),
    \end{equation}
where $F(m,p)$ is a 3D factor depending on the shape of the FS.

   For a 3D cubic lattice for the half-filled band the equation for the FS is 
of the form

    \begin{equation}
    \label{faktor}
    \cos(ak_{F_{x}})+\cos(ak_{F_{y}})+\cos(ak_{F_{z}})=1
    \end{equation}
and we get the equation for $F(m,p)$ 

    \begin{eqnarray}
    F(m,p)= \sin(\arccos(1-\cos (ak_{F_{y}}(m))-\cos (ak_{F_{z}}(p))))\times 
    \nonumber \\ \times \cos(lN\arccos(1-\cos(ak_{F_{y}}(m))-\cos(ak_{F_{z}}
    (p)))).
    \end{eqnarray}
The FS is then an octahedron with the diagonals along $k_{x}, k_{y}$ and $k_{z}
$ axes.

   For the filling factor much less than a half we can expand the cosines in 
relation (\ref{faktor}) for small $k$-values and we obtain

    \begin{equation}
    1-\frac{(ak_{F_{x}})^{2}}{2}+1-\frac{(ak_{F_{y}})^{2}}{2}+1-\frac{(ak_{F_{z}
    })^{2}}{2}=1
    \end{equation}
and $F(m,p)$ takes a form

    \begin{eqnarray}
    F(m,p)=\sin\left(2\sqrt{1-\left(\frac{ak_{F_{y}}(m)}{2}\right)^{2}-\left(
    \frac{ak_{F_{z}}(p)}{2}\right)^{2}}\right)\times \nonumber\\
    \times\cos\left(2lN\sqrt{1-\left(\frac{ak_{F_{y}}(m)}{2}\right)^{2}-\left(
    \frac{ak_{F_{z}}(p)}{2}\right)^{2}}\right).
    \end{eqnarray}
The FS is then spherical.
   
   For a body-centered tetragonal lattice and the half-filled band the equation
for the FS takes a form

    \begin{equation}
    \cos\left(\frac{ak_{F_{x}}}{2}\right)\cos\left(\frac{ak_{F_{y}}}{2}\right)
    \cos\left(\frac{ck_{F_{z}}}{2}\right)=1
    \end{equation}
and $F(m,p)$ is

    \begin{eqnarray}
    F(m,p)= \sin\left(2\arccos\frac{1}{\cos\left(\frac{ak_{F_{y}}(m)}{2}\right)
    \cos \left(\frac{ck_{F_{z}}(p)}{2}\right)}\right)\times \nonumber \\
    \times \cos\left(2lN\arccos\frac{1}{\cos\left(\frac{ak_{F_{y}}(m)}
    {2}\right)\cos\left(\frac{ck_{F_{z}}(p)}{2}\right)}\right).
    \end{eqnarray}
The FS is then a cuboid.

   For a face-centered lattice we have the FS at half filled band

    \begin{equation}
    \cos\left(\frac{ak_{F_{x}}}{2}\right)\cos\left(\frac{ak_{F_{y}}}{2}\right)
    +\cos\left(\frac{ak_{F_{x}}}{2}\right)\cos\left(\frac{ak_{F_{z}}}{2}\right)
    +\cos\left(\frac{ak_{F_{y}}}{2}\right)\cos\left(\frac{ak_{F_{z}}}{2}\right)
    =1
    \end{equation}
and we obtain $F(m,p)$ of the form

    \begin{eqnarray}
    F(m,p)= \sin\left(2\arccos\frac{1-\cos\left(\frac{ak_{F_{y}}(m)}{2}\right)
    \cos\left(\frac{ak_{F_{z}}(p)}{2}\right)}{\cos\left(\frac{ak_{F_{y}}(m)}{2}
    \right)+\cos\left(\frac{ak_{F_{z}}(p)}{2}\right)}\right)\times \nonumber \\
    \times \cos\left(2lN\arccos\frac{1-\cos\left(\frac{ak_{F_{y}}(m)}{2}
    \right)\cos\left(\frac{ak_{F_{z}}(p)}{2}\right)}{\cos\left(\frac{ak_{F_{y}}
    (m)}{2}\right)+\cos\left(\frac{ak_{F_{z}}(p)}{2}\right)}\right).
    \end{eqnarray}
The FS is then a cube.

   Persistent currents can be both paramagnetic and diamagnetic depending on 
the sample dimensions and on the Fermi energy. Paramagnetic currents has been 
studied in details in \cite{Stebelski}, in this paper we will discuss mainly 
the diamagnetic solutions.

   In Fig.2 \marginpar{Insert Fig.2} we present diamagnetic currents for 
different 3D lattices.

   All the above considerations show that the current in a multichannel 
cylinder depends on the strength of the interchannel correlations and 
therefore on the shape of the FS.

   In the case of the spherical (circular) FS the channel currents add almost 
without phase correlation and the total current is small, whereas for the FS 
being a cuboid (square) with rounded corners the correlation is almost perfect 
and the total current is large.

   It follows from (\ref{ost}) and (\ref{chi}) that persistent currents are
carried by coherent electrons. Thus the shape of the FS determines the
density of coherent electrons $n_c$ in the sample and it increases with 
increasing the curvature of the FS.

\section{Self-sustaining currents}

Finally we discuss the possibility of spontaneous self-sustaining currents
i.e. those which flow without any external field.

   The magnetic flux which drives the persistent current given by equation 
(\ref{prad}) is the sum of the  externally applied flux and the flux from the 
currents itself. The inductance coefficient ${\cal L}$ depends on the sample 
geometry and can be large for a cylinder geometry.

   Equations (\ref{strumien}) and (\ref{prad}) form two selfconsistent 
equations for the current and it raises the possibility of a spontaneous 
self-sustaining current at the external flux zero. The possibility of the 
finite spontaneous current depends again crucially on the shape of the FS.

   The phenomenon of the self-sustaining currents is a collective effect and it
requires many electrons to support the current at $\phi_{e}=0.$ Therefore for
a 2D cylinder with $d\sim 1\AA $ we do not find any spontaneous current
solutions. To get it we have to consider the cylinder made of a set of 2D
concentric cylindrical sheets or a 3D cylinder (in both cases we keep $d \ll 
R$).

   The graphical solutions of the self-cosistent equations (\ref{strumien}) and
(\ref{prad}) at $\phi_{e}=0$ are presented in figs. 3, 4 for different shapes
of the FS. The intersections of the two curves marked by circles give the 
values of self-sustaining currents. The temperature at which the transition to 
the state with such currents occurs is denoted by $T_{c}$.

   For diamagnetic currents presented in Fig.3 \marginpar{Insert Fig.3} and 
Fig.4 \marginpar{Insert Fig.4} the self-sustaining currents correspond to 
flux trapping in the cylinder - the phenomenon known in superconductivity.

   It follows from our considerations that the self-sustaining solutions are
obtained only in samples where the FS has flat regions. For spherical and
nearly spherical FS we do not obtain it because the density of coherent
electrons is too small.

   There are two reasons for the presence of coherent electrons in mesoscopic
samples. At first the QSEG hampers the scattering at low temperatures. At 
second as will be shown below, the selfconsistent flux $\phi_{I}$ increases 
coherence in the sample.

   To be mostly transparent let us consider a cylinder made of a set of 
quasi-one dimensional mesoscopic rings stacked along certain axis for the 
filling factor much less than 1. The energy spectrum (cf.(\ref{nowae})) of a 
single ring is given by the formula

    \begin{equation}
    \label{14.1}
    {\cal E}_{s}=\frac{2\hbar^{2}\pi^{2}}{m_{e}L_{x}^{2}}(s-\phi^{'})\;\; s=0,
    \pm 1,\ldots\;\; \phi ^{'}=\phi_{e}^{'}+\frac{{\cal L}I}{\phi_{0}}.
    \end{equation}
Assuming that each ring posses an odd number of electrons we can calculate
the energy gap at the FS, $\Delta \equiv {\cal E}_{s_{F}+1}-{\cal E}_{s_{F}}$.
We find

    \begin{equation}
    \label{14.2}
    \Delta =\Delta_{0} \left( 1- 2\phi_{e}^{'}+2\frac{{\cal L}\mid I \mid}
    {\phi_{0}}\right),
    \end{equation}
where $\Delta_{0}=\frac{h^{2}}{2m_{e}L_{x}^{2}}N$ is the QSEG.

   We see that $\Delta$ contains a term $\Delta_{d}$:

    \begin{equation}
    \label{14.3}
    \Delta_{d}\equiv\Delta_{0}\frac{{\cal L}\mid I \mid}{\phi_{0}},
    \end{equation}
coming from the magnetostatic interactions among electrons - $\Delta_{d}$ is 
the dynamic part of the energy gap, which has to be calculated in a 
self-consistent way.

   If we calculate the energy of electrons in the cylinder made of a set of M
rings for $\phi^{'}<\frac{1}{2}$ we find 

    \begin{equation}
    \label{14.4}
    E(\phi^{'})=M\frac{2\hbar^{2}\pi^{2}}{m_{e}L_{x}^{2}}\sum_{s=0,\pm 1}^{\pm 
    s_{F}}(s-\phi^{'})^{2}<E(\phi_{e}^{'}),
    \end{equation}
because $\phi^{'}=\phi_{e}^{'}+\frac{{\cal L}I}{\phi_{0}}<\phi_{e}^{'}$ as the 
current is diamagnetic for $\phi^{'}<\frac{1}{2}$.

   The gain in the energy in equation (\ref{14.4}) may be called the 
condensation energy due to orbital magnetic interactions.

   In case of an even number of electrons in each ring, spontaneous flux 
$\phi_{sp}$ is created in a system \cite{Szopa}. $\phi_{sp}$ shifts the 
energy levels and as a result a gap appears $\Delta ={\cal E}_{-s_{F}}-{\cal 
E}_{s_{F}}$.

One finds

    \begin{equation}
    \Delta =\Delta_{d}=\Delta_{0}\cdot 2\phi_{sp}^{'},\;\; \phi _{sp}={\cal L}
    I_{sp}, 
    \end{equation}
$I_{sp}$ is the spontaneous current. This gap is dynamic, because it results 
from the collective action of all electrons which produces spontaneous flux in
order to minimize the energy.

   In both cases the flux coming from the currents produces a dynamic gap and
therefore increases coherence of electrons. This mechanism of gaining the
energy is valid only in mesoscopic systems as $\Delta_{d}\rightarrow 0$ for 
$\Delta_{0}\rightarrow 0$. Similar considerations can be performed for
electrons moving on a cylinder with the conclusion that the orbital magnetic
interactions (taken here in the MFA) increase coherence.

   It can easily be seen from equation (\ref{11.4}) that increase of an energy 
gap results in a further depression of $\chi_{p}$ because it decreases both 
terms in the formula for paramagnetic susceptibility thus enhancing the 
coherent response of the system.

This phenomenon is analogous to the reduction of the paramagnetic 
susceptibility in superconductors and in organic molecules due to pair 
correlation \cite{Kresin}.

\section{Conclusions}

   It is already well established both theoretically \cite{Cheung} and
experimentally \cite{Chandrasekhar} that persistent currents which "never 
decay" can flow in mesoscopic metallic or semiconducting rings being a 
manifestation of quantum coherence. Such currents were previously attributed 
solely to superconductors.

   The question arises what are the similarities and differences between the
properties of mesoscopic systems and superconductors \cite{Szopa,Tsiper}. 
Superconductivity is a collective phenomenon and follows from an attractive 
interaction. To compare the two phenomena we considered electrons interacting 
via the orbital magnetic long range interaction and moving on a thin-walled 
hollow cylinder. The interaction, taken here in the MFA, means that each 
electron, besides the external flux $\phi_{e}$ feels a magnetic flux $\phi_{I}$
coming from the currents. The selfinductance of the one-dimensional loop is 
negligible \cite{Trivedi}, however, it can be substantial for a system of 
cylindrical geometry.

   We calculated the frequency dependent conductivity and the electromagnetic
kernel. The conductivity is strongly related to the presence of persistent
currents. We have shown that 

    \begin{equation}
    \label{17.1}
    \lim_{w \rightarrow 0}\lim_{\underline{q}\rightarrow 0}K(\underline{q},w)
    =-c\frac{\partial I}{\partial \phi }
    \end{equation}
The finite limit of the kernel is equivalent to the infinite conductivity. It
is absent in macroscopic normal metals and present in superconductors. Thus
although both the elastic and inelastic scattering was taken into account, part
of electrons ($\equiv n_{c}$) is in a coherent state and moves without 
dissipation. This feature is connected with the multiply connected structure of
our cylinder where a change of phase produced by the magnetic flux can not be 
removed by a gauge transformation and results in an equilibrium current $I$ 
which is persistent at low temperatures.

   In a one-dimensional mesoscopic ring the current $I$ is simply a diamagnetic
or paramagnetic reaction to the external flux $\phi_{e}$ and must vanish at 
$\phi_{e}=0$ \cite{Trivedi}. The situation looks different in a multichannel 
cylinder. We have shown that the self-sustaining, persistent currents can 
exist at $\phi_{e}=0$ in relatively clean samples (ballistic regime) and for 
the FS which are sufficiently flat. The self-sustaining solutions correspond to
"orbital ferromagnetism" for paramagnetic currents (this case was discussed in
\cite{Stebelski}) and to flux trapping for diamagnetic currents - a feature 
characteristic of superconductivity.

   The coherent response and related quantities depend on the sample 
dimensions and on the geometry of the FS. We have shown that the magnitude of a
persistent current and of a dynamic and static response function depend on the 
phase correlation of currents from different channels. This correlation 
increases with increasing the curvature of the FS. The Fermi Surfaces with flat
regions follow in a natural way in the tight-binding model.

   The coherent response of a mesoscopic hollow cylinder follows from two
reasons. At first because of finite size the energy spectrum is discrete and
QSEG causes that the diamagnetic and paramagnetic parts of susceptibility fail 
to cancel - this is connected with the single electron properties. At second 
the orbital magnetic interaction among electrons leads to appearance of the 
dynamic gap and it further increases coherence. It manifests e.g. in further 
decrease of the paramagnetic susceptibility and can lead to the coherent 
collective phenomena such as "orbital ferromagnetism" or flux trapping.

   It follows from our considerations that a mesoscopic cylinder can be in
general described by the two-fluid model. The coherent behaviour is determined
here by the interplay between finite size effects and the correlations
coming from the orbital magnetic interaction. The energy levels are periodic
functions of the flux with period $\phi_{0}$ and the minima of the total energy
determine stable values of flux contained in the cylinder. We will discuss it 
at related problems in more detail in a subsequent paper.

\section{Acknowledgments}
   We thank Dr. M.Szopa for useful discussions.
   Work was supported by Grant KBN PB 1108/P03/95/08.

\newpage
\section*{Figure captions}
Figure 1. Persistent currents in $I_{0}$ units (where $I_{0}=\frac{2et}{N\hbar}
\sin(k_{F}a))$ vs flux for different shapes of the 2D Fermi surfaces for 
$T=1.21 K, E_{F}= 7 eV$ and $\frac{L_{x}}{L_{y}}=3$. In the inserted figure the
shapes of the Fermi surfaces are shown. \vspace{0.5cm}\\
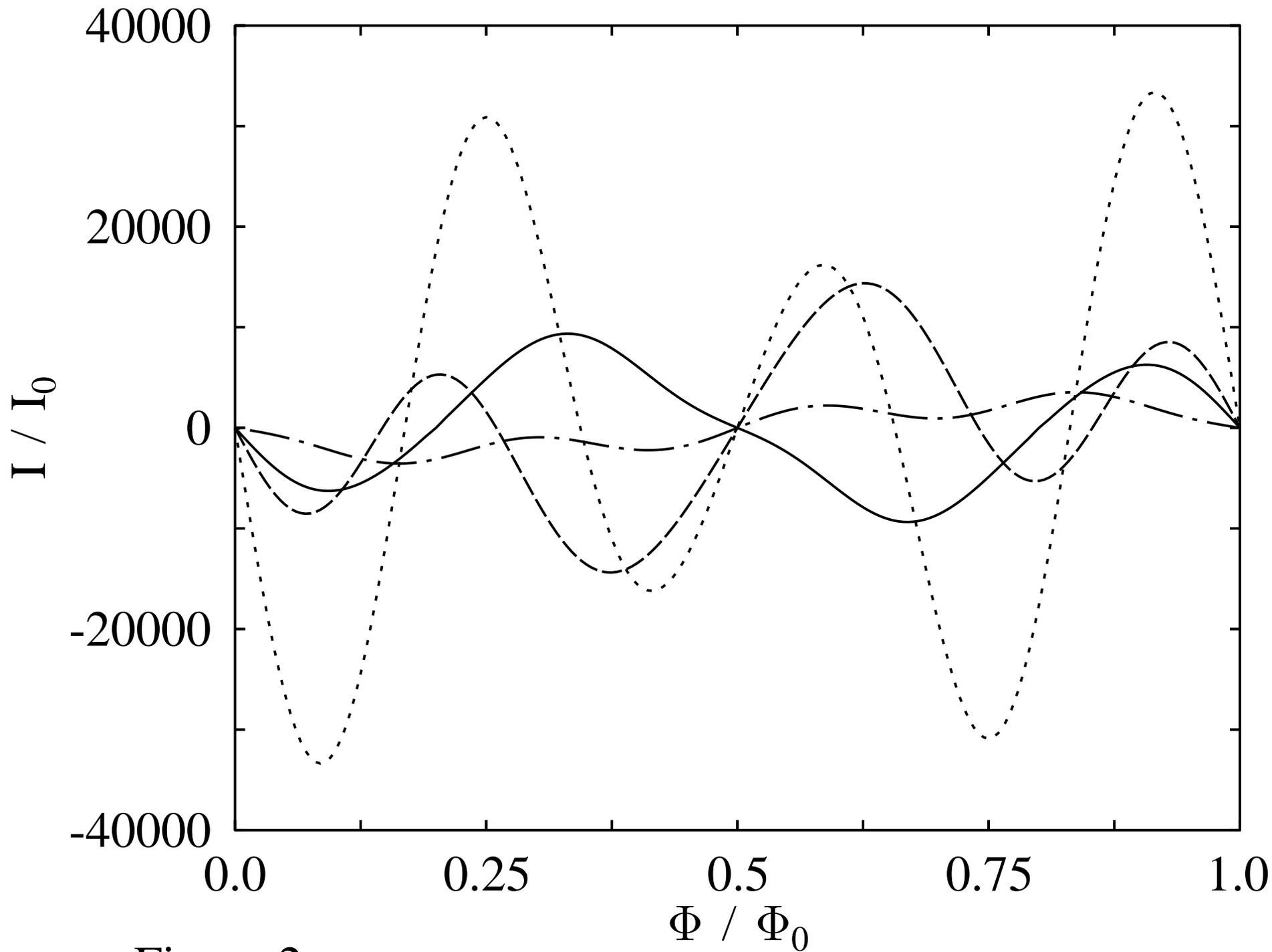
Figure 2. Diamagnetic persistent currents vs flux for different 3D lattices and
slightly less than half filled band: cubic (solid line), body-centered 
tetragonal (dotted line), face-centered (dashed line) and cubic for filling 
factor much less than 1 (dash-dotted line) for $T=1.21 K, E_{F}= 7$ eV and $L_
{x}=3.14 \mu m, L_{y}=0.98 \mu m, L_{z}=0.01 \mu m$. \vspace{0.5cm}\\
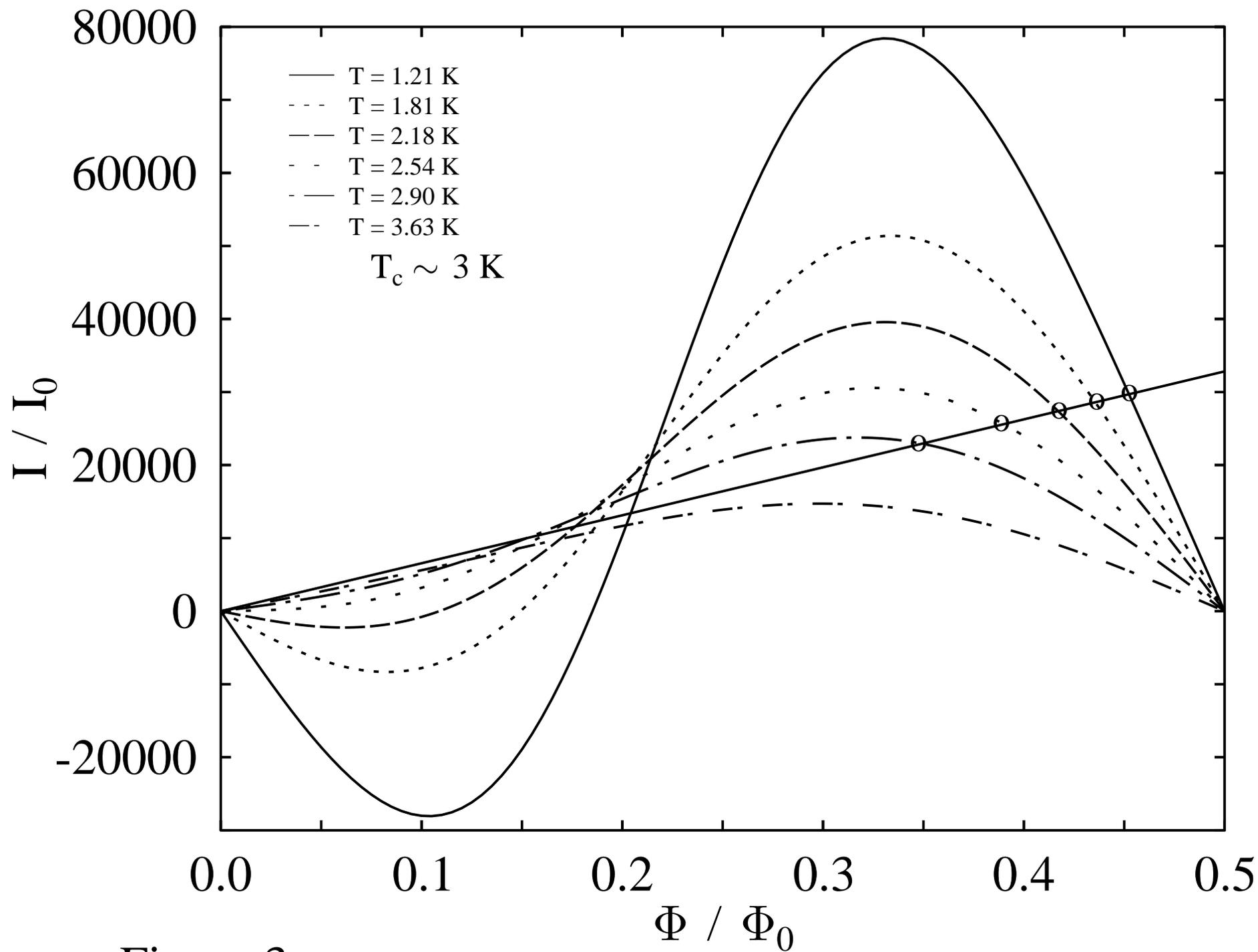
Figure 3. The graphical solution of a set of self-consistent eqs. (\ref{prad})
and (\ref{strumien}) for different temperatures and for body-centered 
tetragonal lattice for $E_{F}=8$ eV and $L_{x}= 3.14 \mu m, L_{y}=0.94 \mu m, 
L_{z}=0.02 \mu m$. The nonzero crossings of the straight line (\ref{strumien})
with the current-flux characteristic (\ref{prad}) denoted by circles correspond
to flux trapped in mesoscopic cylinder. \vspace{0.5cm}\\
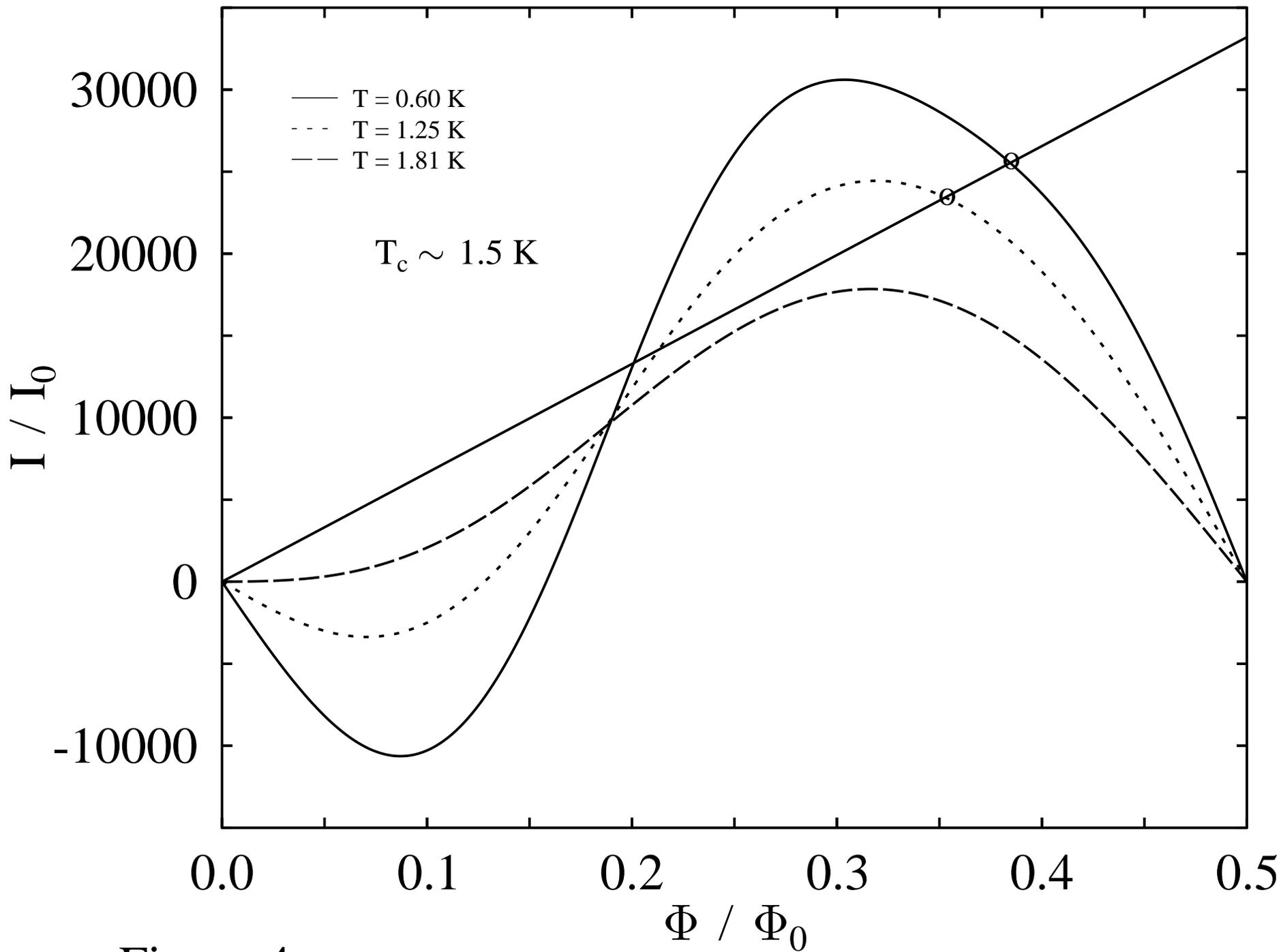
Figure 4. The graphical solution of a set of self-consistent eqs. (\ref{prad})
and (\ref{strumien}) for different temperatures and for face-centered lattice. 
for $E_{F}=6$ eV and $L_{x}=3.14 \mu m, L_{y}=0.96 \mu m, L_{z}=0.03 \mu m$.
The nonzero crossings of the straight line (\ref{strumien}) with the
current-flux characteristic (\ref{prad}) denoted by circles correspond to flux
trapped in mesoscopic cylinder. \vspace{0.5cm}\\
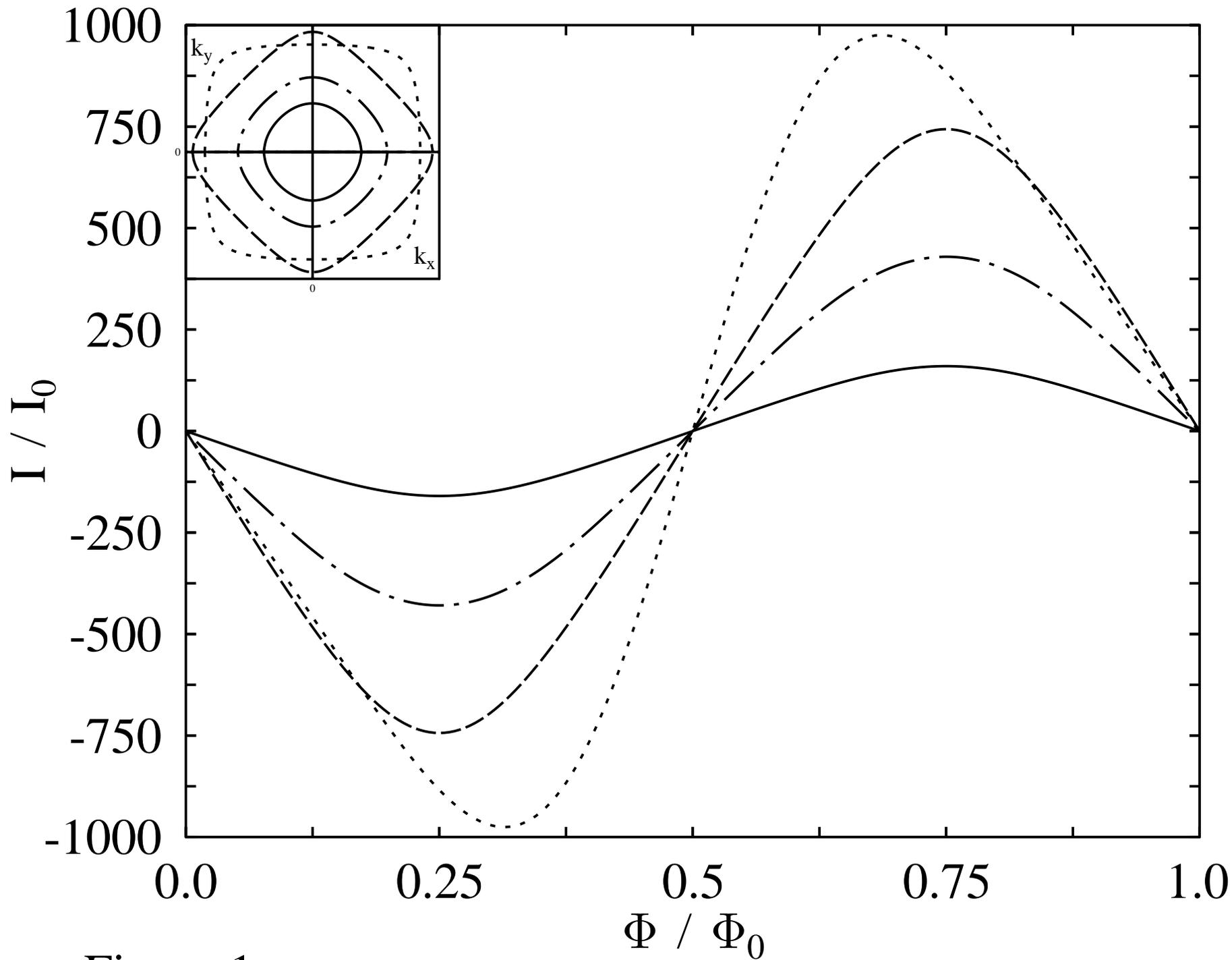
\end{document}